\documentclass[
 aip,
 apl,
 amsmath,amssymb,
 preprint
]{revtex4-1}

\usepackage{graphicx}
\usepackage{dcolumn}
\usepackage{bm}
\usepackage{siunitx}

\usepackage[utf8]{inputenc}
\usepackage[T1]{fontenc}
\usepackage{mathptmx}
\usepackage{etoolbox}
\usepackage{comment}

\makeatletter
\def\@email#1#2{%
 \endgroup
 \patchcmd{\titleblock@produce}
  {\frontmatter@RRAPformat}
  {\frontmatter@RRAPformat{\produce@RRAP{*#1\href{mailto:#2}{#2}}}\frontmatter@RRAPformat}
  {}{}
}
\makeatother

\newcommand\EJ{E_\mathrm{J}}
\newcommand\signal{a}
\newcommand\idler{b}
\newcommand\signalIn{s}

\newcommand\omegaJ{\omega_\mathrm{J}}
\newcommand\omegaS{\omega_\mathrm{\signal}}
\newcommand\omegaI{\omega_\mathrm{\idler}}
\newcommand\omegaSI{\omega_\mathrm{\signal, \idler}}
\newcommand\omegaSin{\omega_\mathrm{\signalIn}}

\newcommand\phiS{\varphi_\mathrm{\signal}}
\newcommand\phiI{\varphi_\mathrm{\idler}}
\newcommand\phiSI{\varphi_\mathrm{\signal,\idler}}
\newcommand\width{\Gamma}
\newcommand\widthS{\width_\mathrm{\signal}}
\newcommand\widthI{\width_\mathrm{\idler}}
\newcommand\widthSI{\width_\mathrm{\signal,\idler}}
\newcommand\tuning\eta
\newcommand\tuningS{\tuning_\mathrm{\signal}}
\newcommand\tuningI{\tuning_\mathrm{\idler}}
\newcommand\detuning\Delta
\newcommand\detuningS{\detuning_\mathrm{\signal}}
\newcommand\detuningI{\detuning_\mathrm{\idler}}

\newcommand\ZSI{Z_\mathrm{\signal,\idler}}

\newcommand\gain{S_\mathrm{\signal\signal}}
\newcommand\maxgain{\gain^{\mathrm{max}}}

\begin{document}

\preprint{AIP/123-QED}

\title[Influence of bias voltage noise on the ICTA]{Influence of bias voltage noise on the \\ Inelastic Cooper-Pair Tunneling Amplifier (ICTA)}

\author{U. Martel}
\affiliation{ 
Institut Quantique, Université de Sherbrooke, Sherbrooke, Québec, Canada J1K 2R1}
\author{R. Albert}
\author{F. Blanchet}
 \affiliation{ 
Univ. Grenoble Alpes, CEA, INAC-PHELIQS, F-38000 Grenoble, France}
\author{J. Griesmar}
\author{G. Ouellet}
\author{H. Therrien}
\author{N. Nehra}
\author{N. Bourlet}
 \affiliation{ 
Institut Quantique, Université de Sherbrooke, Sherbrooke, Québec, Canada J1K 2R1}
\author{A. Peugeot}
 \affiliation{CNRS, Laboratoire de Physique, Ecole Normale Supérieure de Lyon, Lyon F-69342, France}
\author{M. Hofheinz}
 \email{max.hofheinz@usherbrooke.ca}
 \affiliation{ 
Institut Quantique, Université de Sherbrooke, Sherbrooke, Québec, Canada J1K 2R1}

\date{\today}

\begin{abstract}
We experimentally show that the Inelastic Cooper-Pair Tunneling Amplifier (ICTA), implementing a DC-powered parametric amplification scheme, can achieve gain and noise performance similar to that of AC-powered Josephson parametric amplifiers. Using experimental data and simulations, we show that the ICTA has near-quantum-limited noise as long as low-frequency voltage noise, expressed as broadening of the Josephson frequency line, is narrower than the amplification bandwidth. We observe a gain of 20 dB across a 11\,MHz bandwidth with noise below 1.7 times the quantum limit when the full width at half maximum of the Josephson-frequency linewidth is 5.6 MHz. 
\end{abstract}

\maketitle

Quantum-limited linear amplifiers, with input-referred noise power spectral density of one photon at high gain are a versatile means to access microwave quantum systems. In particular, they are an essential ingredient for readout of state-of-the-art superconducting quantum processors\cite{Kaufman2023} based on circuit quantum electrodynamics\cite{Blais2021}, and have enabled progress in many other fields\cite{Aumentado2020}. Quantum-limited microwave amplification has so far only been achieved with parametric amplifiers where a pump tone is downconverted into a signal and an additional idler mode via a nonlinearity such as a Josephson junction or kinetic inductance. Spontaneous downconversion due to the vacuum fluctuations of the signal and idler modes results in the unavoidable quantum-limited noise of these amplifiers. Parametric amplifiers have been built in various configurations, such as resonant amplifiers \cite{Bergeal2010a, Mutus2013, Roy2015, Frattini2018} working in reflection and achieving the lowest noise, or traveling-wave amplifiers (TWPAs) \cite{HoEom2012, Yaakobi2013, Bockstiegel2014, Macklin2015, Goldstein2020, Esposito2021} achieving best bandwidth. 

However, these parametric amplifiers are powered by a strong AC pump tone which requires hardware overhead for generation and routing. In addition, the pump tone may perturb the device under test or saturate the subsequent readout chain. Filtering this tone usually requires additional hardware overhead in the form of circulators, diplexers, or filters that cause loss in the signal path, degrading the system noise performance. DC-powered superconducting amplifiers using Josephson junctions, such as the SLUG \cite{Hover2012} and the SJA \cite{Laehteenmaeki2012}, do away with the AC pump, but suffer from higher noise because they lack a dedicated cool idler mode. Instead, in these devices, a resistive shunt across the junction, required to stabilize the working point, also functions as idler. However, in stabilizing the working point, this resistor absorbs DC current, becomes hot, and causes excess noise. In addition, its frequency is ill defined so that parasitic frequency conversion (beam-splitter) interactions may degrade performance. 

We have previously demonstrated a scheme similar to a Josephson parametric amplifier with a dedicated idler-mode resonator, where the AC pump tone is replaced by a voltage-biased Josephson junction\cite{Jebari2018}. In this configuration, Cooper pairs tunnel inelastically through a Josephson junction and provide energy to the signal and idler fields, similar to pump photons in parametric amplifiers. This DC-powered Inelastic-Cooper-pair-Tunneling Amplifier (ICTA) allows, in principle, to reach the quantum limit. Nevertheless, with the first implementation of this device \cite{Jebari2018} we were not able to reach gains higher than approximately 10 dB while keeping noise within a factor 2 of the quantum limit. 

Here we show that the limiting factor for higher gain is low-frequency noise on the voltage bias, which is equivalent to phase noise of the pump tone in JPAs. The general idea is that the window of optimal bias voltage, expressed as Josephson frequency, is closely related to the bandwidth of the amplifier and becomes narrower with increasing gain. When it becomes narrower than the broadening of the Josephson frequency due to low-frequency voltage noise, instantaneous gain below the optimal gain becomes increasingly likely. The average power gain is then $\langle \gain\rangle^2$, where $\gain$ is the instantaneous amplitude gain. If the amplifier is quantum limited at each instant, the average output noise is $\langle|\gain|^2-1\rangle$. As $\langle \gain\rangle^2 < \langle|\gain|^2\rangle$, unless $\gain$ is constant, the voltage fluctuations reduce gain more strongly than noise, degrading the input-referred noise. Therefore, voltage noise limits the maximal gain that can be achieved while maintaining noise close to the quantum limit. 

We first derive the relation of gain, bandwidth and optimal bias-voltage range of the ICTA and then show that they explain the experimental amplifier noise for different gain and bias-voltage noise. 

\begin{figure}[t]
    \includegraphics[width=0.48\textwidth]{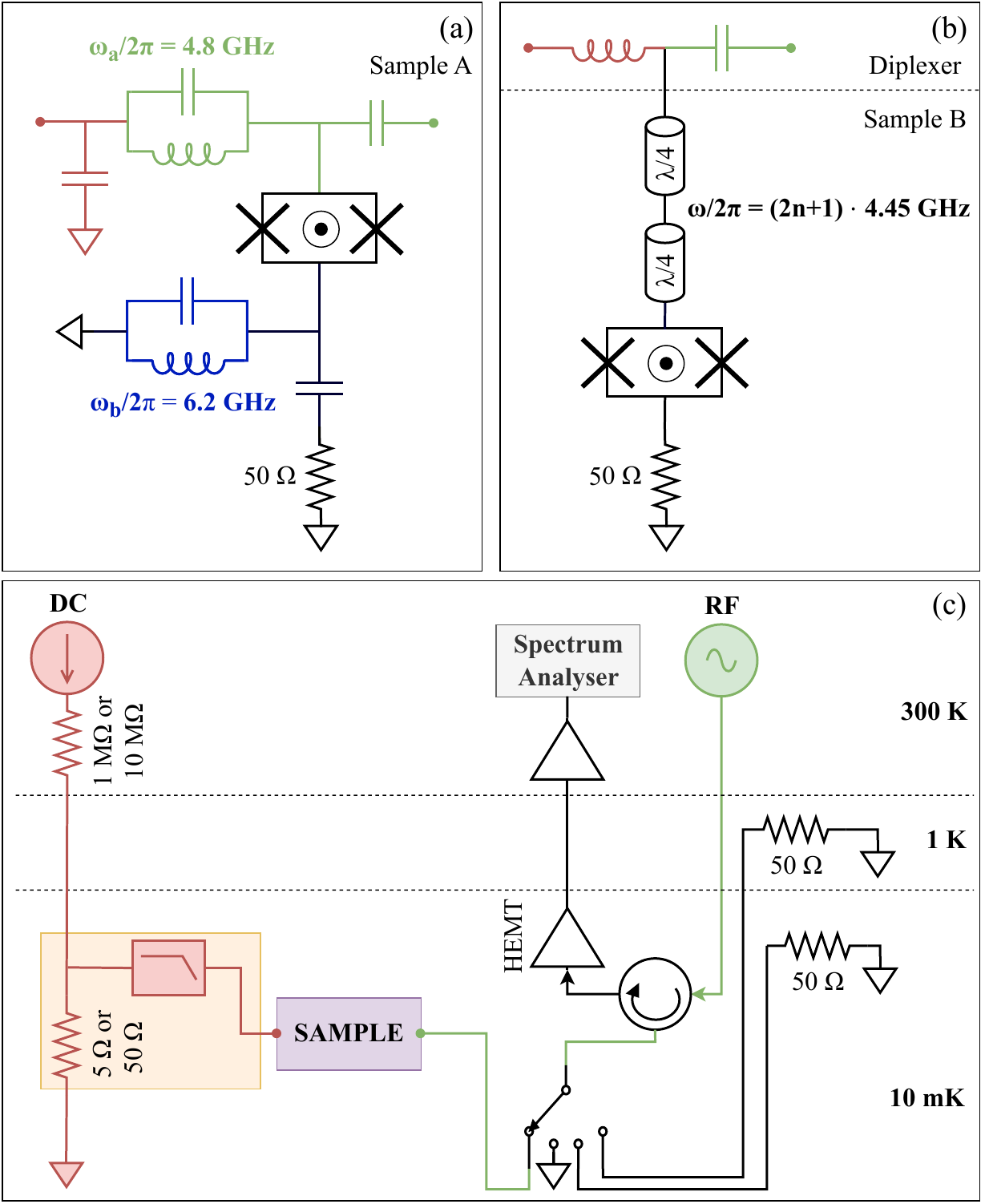}
    \caption{\label{fig:setup} Setup and sample schematics. (a) Sample A, with on-chip bias-tee. (b) Sample B, with off-chip diplexer. (c) Measurement chain. The orange box is the biasing circuit (see supplementary material for details). The 1\,M$\Omega$ resistor is used with the 5\,$\Omega$ biasing circuit and the 10\,M$\Omega$ resistor is used with the 50\,$\Omega$ biasing circuit.
    }
\end{figure}

The ICTA circuits we consider are shown in Fig.~\ref{fig:setup}a and b and consist of a Josephson junction with Josephson energy $\EJ$ and phase $\phi$, in series with a voltage bias $V$ and two resonators of frequencies $\omegaS$ and $\omegaI$.  The corresponding ICTA Hamiltonian is \cite{JebariThese, MartelThese}
\begin{align}
    H = \hbar{\omegaS}{\signal^\dagger \signal} + \hbar{\omegaI}{\idler^\dagger \idler} - \EJ\cos(\phi)
    \label{eq:Ham}
\end{align}
with
\begin{align*}
    \phi = {\omegaJ} t + {\phiS} ({\signal^\dagger + \signal}) + {\phiI} ({\idler^\dagger + \idler})
\end{align*}
where $\omegaJ = {2eV}/\hbar$ is the Josephson frequency, and $\phiSI = \sqrt{\pi \frac{4e^2}{h} \ZSI}$ are the zero-point fluctuations of phase of modes $\omegaSI$, with $\ZSI$ their characteristic impedances.

Supposing small signals and low $\phiSI$ we can develop the cosine term, 
\begin{align}
    -\EJ \cos(\phi) = -\frac{\EJ}{2}[e^{i{\omegaJ} t}e^{i{\phiS} ({\signal^\dagger + \signal})}e^{i{\phiI} ({\idler^\dagger + \idler})} + \text{h.c.}] \, ,
    \label{eq:H1_ini}
\end{align}
 to second order in $\phiS\signal$ and $\phiI\idler$. 

Assuming low Josephson energy, i.e.\ $\phiSI\EJ/\hbar$ is small compared to the mode frequencies, we then perform a rotating wave approximation at $\omegaJ \approx \omegaS + \omegaI$. Together, these approximations yield the well-known parametric amplifier Hamiltonian
\begin{align}
    H = \hbar{\omegaS}{\signal^\dagger \signal} + \hbar{\omegaI}{\idler^\dagger \idler} + \hbar \lambda ({\signal^\dagger}{\idler^\dagger} e^{-i{\omegaJ} t} + \text{h.c})\,,
    \label{eq:Ham_ICTA}
\end{align}
with $\lambda = \frac{\EJ \phiS\phiI}{2\hbar}$. It maps to the JPA Hamiltonian \cite{Abdo2013} with the voltage bias $\omegaJ$ and the Josephson energy $E_J$ in the ICTA playing, respectively, the role of pump frequency and pump amplitude in the JPA.

Using input-output theory to eliminate the cavity modes in the same way as for the JPA \cite{Bergeal2010, Abdo2013, Roy2016}, we can show that
\begin{align}
    \gain = \frac{\tuningS \tuningI + \Xi^2}{\tuningS^{*}\tuningI-\Xi^2}
    \label{eq:gain_det}
\end{align}
where we have defined $\Xi = \frac{2\lambda}{\sqrt{\widthS\widthI}}$ with $\widthS$ and $\widthI$ being the relaxation rates of the resonators, and
\begin{align}
    \tuning_{x} = 1 + 2i \frac{\detuning_{x}}{\width_x}
\end{align}
with $\detuningS = \omegaSin - \omegaS$ the detuning between the signal frequency and the signal mode and $\detuningI = \omegaJ -\omegaSin-\omegaI$ the detuning between the idler frequency $\omegaJ - \omegaSin$ and the idler mode $\omegaI$. The maximum of gain $\gain$, reached for $\detuningS = \detuningI=0$, is then
\begin{align}
\maxgain = \frac{1 + \Xi^2}{1 - \Xi^2}.  
\label{eq:gain_max}
\end{align}
In the absence of voltage noise, under optimal bias condition $\omegaJ = \omegaS + \omegaI$, we have $\detuningS = -\detuningI =: \detuning$ and in the limit of high gain, the gain profile is approximately Lorentzian 
\begin{align}
    \gain \stackrel{\Xi \xrightarrow{} 1^-}{\rightarrow} \frac{\maxgain}{1-i\maxgain\frac{\detuning}{\width}},
    \label{eq:gain_freq}
\end{align}
where $\width^{-1} := \widthS^{-1} + \widthI^{-1}$. It follows that the bandwidth is
\begin{align}
    B_0 = \frac{2\width}{\maxgain}.
    \label{eq:bandwidth}
\end{align}

To address the impact of voltage noise, we instead assume that the input signal is held at maximal gain, i.e. $\detuningS=0$, but the bias voltage changes due to voltage noise. These voltage fluctuations will then only detune the idler from resonance, and $\detuning := \detuningI = \omegaJ-\omegaS-\omegaI$ and $\width := \widthI$ in  Eqs.~(\ref{eq:gain_freq}) and (\ref{eq:bandwidth}). 

Comparing these two scenarios, shows that the amplifier bandwidth (i.e. the signal frequency range where the gain stays within \SI{3}{\decibel} of the maximal gain) and optimal bias-voltage range (i.e. the Josephson frequency range where the gain stays within \SI{3}{\decibel} of the maximum gain) are of the same order of magnitude. For the degenerate case, where signal and idler reside in the same mode, the optimal bias-voltage range is exactly twice the bandwidth. Both widths are inversely proportional to the maximal gain.

The experimental setup is shown in Fig.~\ref{fig:setup}c. A DC bias is applied to the ICTA through a voltage divider and filter (see supplementary material) with output impedance of either \SI{50}{\ohm} or \SI{5}{\ohm}. The voltage-bias noise is dominated by the thermal noise of this output impedance. The different output impedances of the filters, therefore, result in different voltage-bias noise. A SQUID acts as a flux tunable Josephson junction which we use to adjust the gain of the ICTA via an on-chip flux line,  biased by a room temperature voltage source in series with a \SI{5}{\kilo\ohm} resistor and filtered at base temperature using a custom dissipative low-pass filter \cite{Paquette2022} with a cutoff frequency of the order of \SI{200}{\mega\hertz}.

To measure gain, a microwave signal is sent to the sample through an attenuated line, a circulator, a switch and a high-pass filter. The circulator then routes the signal reflected from the device to a cyrogenic HEMT amplifier at 4K. The signal is then further amplified, downconverted and digitized using a custom double-heterodyne receiver, phase locked to the input signal. This setup allows for scalar network analyzer measurements and noise measurements using the same signal pathway.

The switch allows us to also connect the readout chain to a short circuit and two loads, one thermally anchored to the mixing chamber stage of the fridge and one to the still stage at approximately \SI{1}{\kelvin}. These different microwave loads are used for Y-factor calibration of the noise measurement (see supplementary material). 

Fig.~\ref{fig:setup}a and Fig.~\ref{fig:setup}b show the two samples measured in this article (see supplementary material for additional details). On sample A, LC resonators form the signal and idler modes, with center frequencies $\omegaS/2\pi = \SI{4800}{\mega\hertz}$ and $\omegaI/2\pi = \SI{6200}{\mega\hertz}$, widths $\widthS/2\pi = \SI{96}{MHz}$ and $\widthI/2\pi = \SI{226}{MHz}$, and estimated characteristic impedances $\ZSI \approx \SI{400}{\ohm}$.  DC bias and the microwave signal are applied through separate ports, allowing further noise filtering directly on chip (see supplementary material). Sample B uses a degenerate mode for signal and idler formed by the fundamental mode of a $\lambda/4$ resonator with center frequency $\omegaSI/2\pi  = \SI{4450}{\mega\hertz}$, width $\widthSI/2\pi = \SI{185}{\mega\hertz}$ and an estimated resonance (not transmission-line) characteristic impedance of $\ZSI \approx \SI{80}{\ohm}$. DC and microwave signals are connected to the same port and split using an off-chip diplexer (Marki DPXN-M50) with a \SI{50}{\mega\hertz} cross-over frequency. Samples are fabricated using a self-aligned process \cite{Grimm2017} with Nb/Al/AlOx/Nb junctions, two Nb routing layers, separated by SiN dielectric. 

\begin{figure}[t]
    \includegraphics[width=0.49\textwidth]{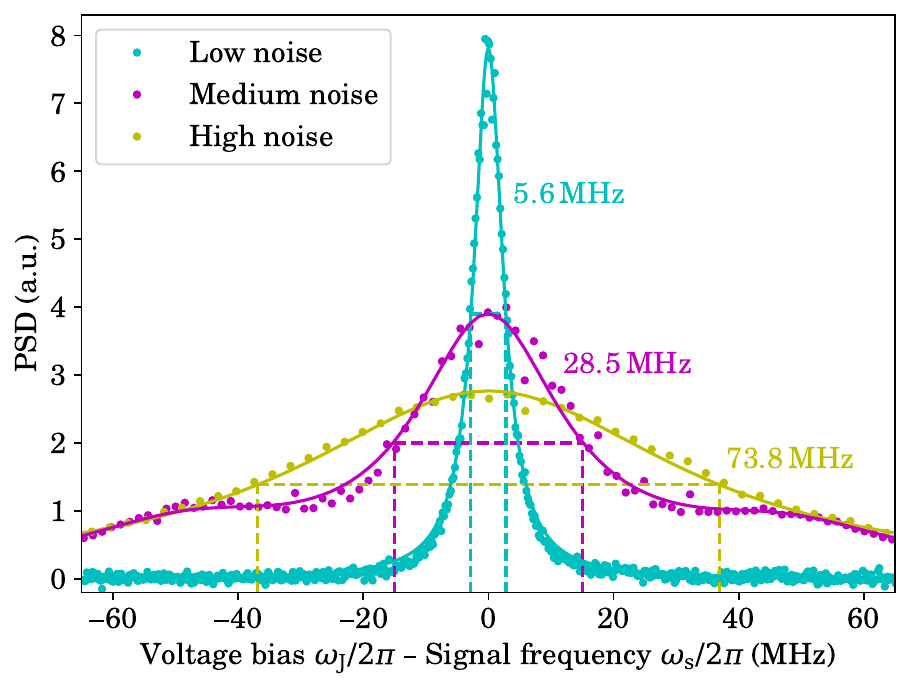}
    \caption{\label{fig:voltage_noise} Josephson frequency distribution (dots) for the three configurations, through measurement of the AC Josephson effect at minimum Josephson energy $\EJ$. Frequency $\omegaSin / 2\pi$ was fixed at 4772\,MHz for sample A (low noise) and 4122\,MHz for Sample B (medium and high noise). The low and high noise curves were each fitted to a single Lorentzian distribution (solid lines), whereas three were needed for the medium noise curve. Dashed lines indicate FWHM of each curve. For the medium noise, the FWHM of the central peak is shown.}
\end{figure}

We thus have three different experimental configurations for voltage noise: low noise with sample A with on-chip filtering combined with the \SI{5}{\ohm} bias circuit, medium noise with sample B without on-chip filtering, combined with the \SI{5}{\ohm} bias circuit, and high noise with Sample B combined with the \SI{50}{\ohm} bias circuit. 
We characterize the resulting voltage-bias distribution using the AC-Josephson effect. We first maximally frustrate the SQUIDs so that spontaneous emission by the devices is minimal and can be described by $P(E)$ theory \cite{Ingold1992,Hofheinz2011}. We then measure the power spectral density emitted by the devices at a fixed frequency $\omega$ and vary the bias voltage, expressed as Josephson frequency $\omegaJ$, around $\omega$, giving the Josephson-frequency  distribution.  If the low-frequency impedance of the Josephson junction bias $Z$ is $\ll \frac{h}{4e^2}$ and flat and if the Josephson energy is low, this distribution is expected to be a Lorentzian \cite{Ingold1991,DiDomenico2010,Hofheinz2011} with a FWHM of 
\begin{equation}
\Delta\omegaJ = 2 k_\mathrm{B} T \frac{4e^2}{\hbar^2} Z(0).
\label{eq:fwhm}
\end{equation}
Fig.~\ref{fig:voltage_noise} shows that for the low-noise configuration and high-noise configuration, the voltage noise distribution is well fitted by the expected Lorentzian shape, with full widths at half maximum of, respectively, $\Delta\omegaJ/2\pi = \SI{5.6}{\mega\hertz}$ and $\Delta\omegaJ/2\pi = \SI{73.8}{\mega\hertz}$, corresponding to effective temperatures of the bias impedances of, respectively, \SI{27.6}{\milli\kelvin} and \SI{36.4}{\milli\kelvin}. In the medium-noise configuration, additional bumps appear in the distribution at the crossover frequency of the diplexer where the bias impedance changes from \SI{5}{\ohm} to \SI{50}{\ohm}, leading to side lobes in the distribution\cite{DiDomenico2010}. A combination of three Lorentzian distributions were used to fit this configuration, resulting in full widths at half maximum (FWHM) of $\Delta\omegaJ/2\pi = \SI{28.5}{\mega\hertz}$ for the central peak and $\Delta\omegaJ/2\pi = \SI{45.8}{\mega\hertz}$ for the two side peaks at $\SI{\pm 48}{\mega\hertz}$. Note that only voltage noise at frequencies below approximately $\Delta \omegaJ/16\ln (2)$ contributes to the width $\Delta \omegaJ$ of the distribution \cite{DiDomenico2010}. As this cutoff is lower than the bandwidth of the amplifier, we may in the following consider fluctuations of $\omegaJ$ to be adiabatic. 

\begin{figure}[b]
   \includegraphics[width=0.49\textwidth]{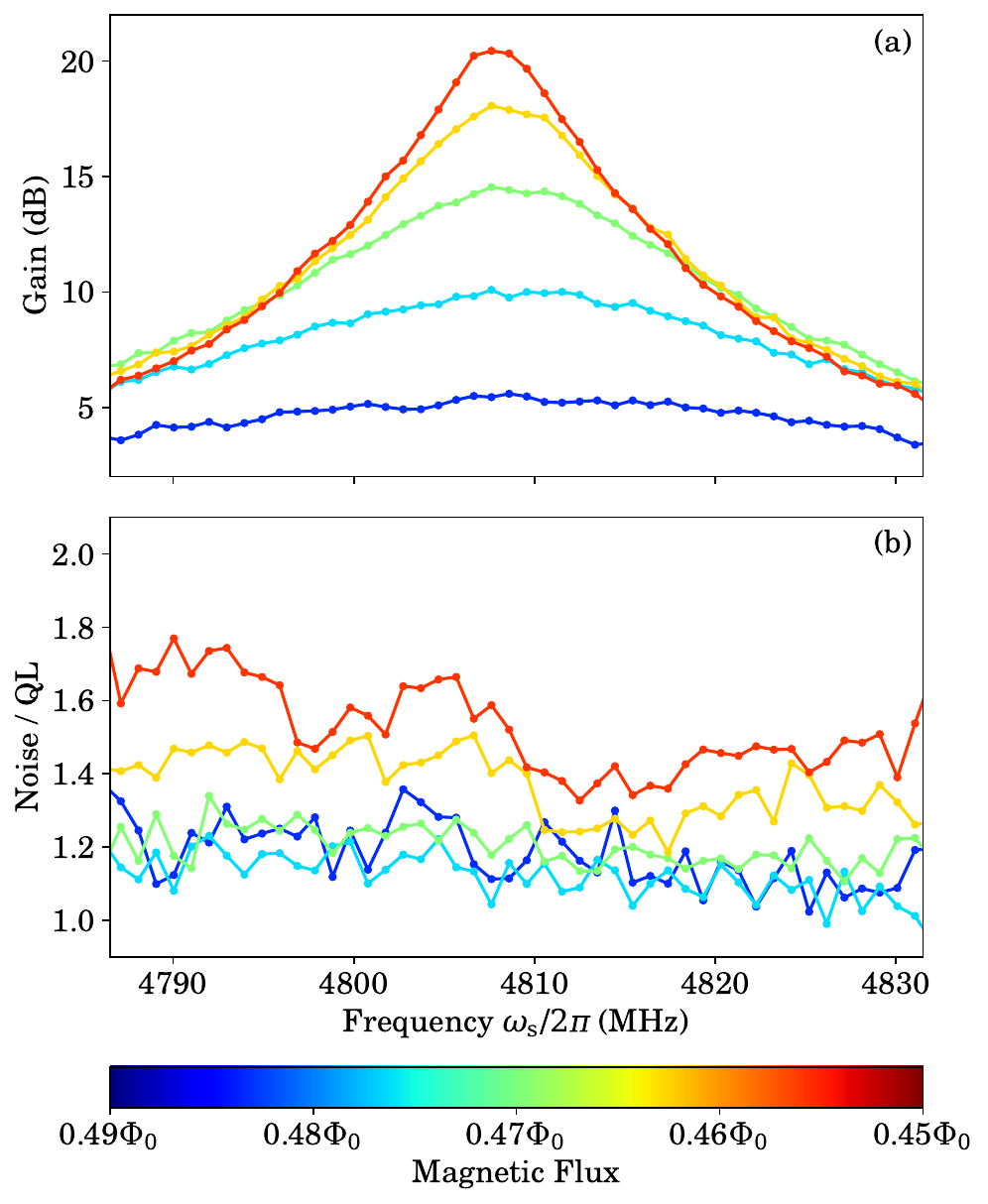}
    \caption{\label{fig:gain_noise} Gain (a) and noise referred to the quantum limit (b) of the low noise configuration (sample A with \SI{5}{\Omega} bias circuit) as a function of magnetic flux and frequency for a voltage bias of $\omegaJ /2\pi = \SI{10991}{\mega\hertz}$, with a idler mode $\omegaI/2\pi = \SI{6181}{\mega\hertz}$. (See supplementary material for details on calibration.)}
\end{figure}

Fig.~\ref{fig:gain_noise} shows typical gain and noise curves for the low-noise configuration at fixed voltage bias and for different Josephson energies tuned via the flux bias to the SQUID. Magnetic flux close to half a flux quantum, corresponding to the lowest Josephson energy $\EJ$, gives the lowest gain and highest bandwidths, as expected from Eq.~(\ref{eq:bandwidth}). For the three lowest gain curves, the noise is virtually the same, between $1$ and $1.3$ times the quantum limit. Increasing the gain, and thereby reducing the bandwidth, slightly increases the noise but it remains below 1.5 and 1.7 times the quantum limit up to \SI{18}{\decibel} and \SI{20.5}{\decibel}, respectively. 

\begin{figure}
  \includegraphics[width=0.49\textwidth]{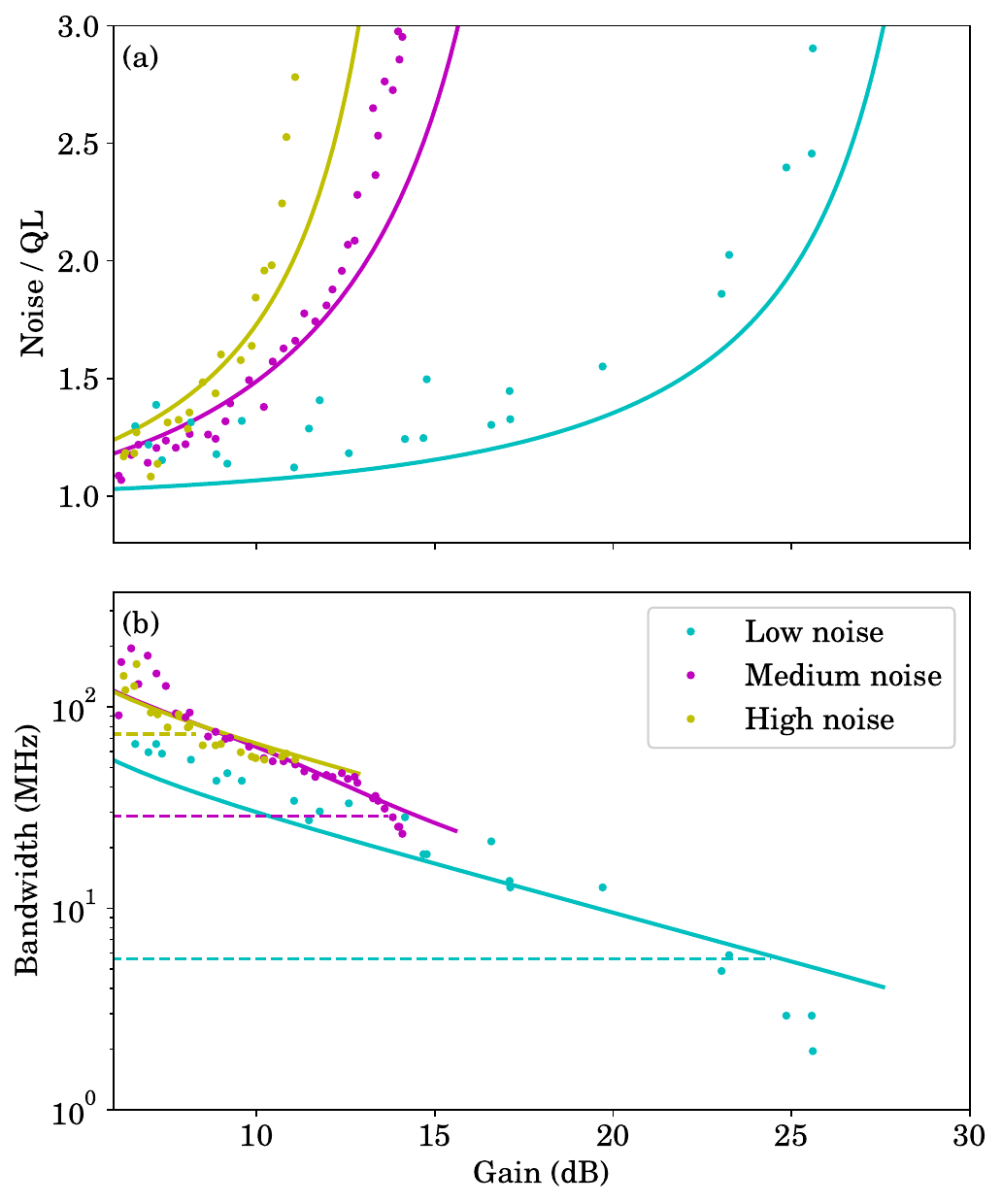}
    \caption{\label{fig:scatter} Noise with respect to the quantum limit (a) and bandwidth (b) as a function of gain, adjusted via effective Josephson energy $\EJ$, and at fixed voltage bias $\omegaJ /2\pi$ at 10952\,MHz for Sample A (low noise) and 8982\,MHz for Sample B (medium and high noise). Signal frequency $\omegaSin /2\pi$ is 4771\,MHz for Sample A and 4540\,MHz for Sample B. Dots are measurements, solid lines correspond to an ideal parametric amplifier following Eq.~(\ref{eq:gain_det}) with quantum limited noise, averaged over the fitted voltage bias distributions (see Fig.~\ref{fig:voltage_noise}), with measured signal and idler mode widths and $\Xi$ as sweep parameter. Dashed lines are the FWHM of the corresponding voltage noise distributions.}
\end{figure}

In Fig.~\ref{fig:scatter} we plot the gain, bandwidth and noise extracted from similar curves for the low noise, medium noise and high noise configurations. Data was taken by varying the Josephson energy while using a fixed voltage bias for each configuration. Like in Fig.~\ref{fig:gain_noise}, we can see in Fig.~\ref{fig:scatter}a that raising the gain also raises the noise. For the same gain, we see lower amplification noise with lower voltage noise. The maximum gain we can achieve while remaining below three times the quantum limit is 11.1\,dB for the highest noise configuration, \SI{14.1}{\decibel} for the medium one, and \SI{25.6}{\decibel} for the lowest one. 

We compare these experimental results with Eq.~(\ref{eq:gain_det}) averaged over the fitted voltage bias distributions determined from the data in Fig.~\ref{fig:voltage_noise}, approximating the voltage bias fluctuations as adiabatic and using $\Xi$ as a variable. These numerical calculations closely follow the experimental data without any fitting parameter, even though they slightly underestimate the noise. The slightly higher noise in the experimental results may be due to losses in the device.

Fig.~\ref{fig:scatter}b shows the relationship between bandwidth and gain. Both experimental and numerical results show the expected constant gain$\times$bandwidth product expected from Eq.~(\ref{eq:gain_freq}). By design, Sample B has resonators with lower Q-factors, resulting in higher gain$ \times$bandwidth product.

Comparing Figs.~\ref{fig:scatter}a and b and the width of the Josephson frequency distribution (dashed lines in Figs.~\ref{fig:scatter}b), we indeed see that the amplifier noise diverges when the bandwidth (which is equal to optimal bias voltage range within a numerical factor) becomes comparable to the width of the Josephson frequency distribution. 

In conclusion, we have investigated the impact of voltage-bias noise on ICTA performance, a noise source that corresponds to pump phase noise in JPAs and is usually negligible there. We have shown experimentally that bias-voltage noise does not degrade ICTA noise as long as the resulting distribution of Josephson frequencies is narrower than the bandwidth of the device (or more precisely the optimal Josephson frequency range), so that voltage noise does not cause gain fluctuations. 
This result implies that improving the bandwidth of the ICTA, e.g.\ via impedance engineering \cite{Roy2015}, promises to improve at the same time noise and robustness of the ICTA. 
Because it is DC-powered and does not require a pump tone, the ICTA then avoids the risk of saturating the amplifier chain and the need of additional isolation, simplifying the use of quantum-limited amplifiers.

\section*{Supplementary Material}

See supplementary material for details on the voltage biasing circuits, calibration,  circuit parameters and an image of the device.

\section*{Author Declarations}

\subsection*{Conflict of Interest}
The authors have no conflicts to disclose.

\subsection*{Author Contributions}

\textbf{Ulrich Martel}: Conceptualisation (supporting), Formal analysis (equal), Investigation (lead), Methodology (equal), Software (supporting), Visualisation (lead), Writing - original draft (equal), Writing - review \& editing (equal).
\textbf{Romain Albert}: Investigation (supporting), Methodology (supporting), Software (supporting), Writing - review \& editing (equal).
\textbf{Florian Blanchet}: Conceptualisation (supporting), Investigation (supporting), Methodology (supporting), Software (lead).
\textbf{Joel Griesmar}: Formal analysis (supporting), Investigation (supporting), Methodology (supporting), Software (supporting), Supervision (supporting).
\textbf{Gabriel Ouellet}: Investigation (supporting), Methodology (supporting), Software (supporting).
\textbf{Hugo Therrien}: Investigation (supporting), Methodology (supporting).
\textbf{Naveen Nehra}: Conceptualisation (supporting), Writing - review \& editing (supporting).
\textbf{Nicolas Bourlet}: Formal analysis (supporting), Investigation (supporting), Methodology (supporting), Supervision (supporting), Writing - review \& editing (supporting).
\textbf{Ambroise Peugeot}: Formal analysis (supporting).
\textbf{Max Hofheinz}: Conceptualisation (lead), Formal analysis (equal), Investigation (supporting), Methodology (equal), Software (supporting), Writing - original draft (equal), Writing - review \& editing (equal), Supervision (lead), Funding acquisition (lead), Validation (lead).

\begin{acknowledgments}
This work was supported by the Natural Sciences and Engineering Research Council of Canada, the Canada First Research Excellence Fund, the European Union (ERC
Starting Grant No. 278203 WiQOJo), and the French Agence Nationale de la Recherche (grant JosePhSCharLi ANR-16-CE92-0033).
\end{acknowledgments}

\section*{Data Availability Statement}

The data that support the findings of this study are available from the corresponding author upon reasonable request.

\bibliography{article_biblio}

\end{document}


\setcounter{figure}{0}
\renewcommand{\figurename}{Fig.}
\renewcommand{\thefigure}{S\arabic{figure}}
\renewcommand{\theequation}{S\arabic{equation}}
\renewcommand{\thetable}{S\arabic{table}}

\preprint{AIP/123-QED}

\title[Influence of bias voltage noise on the ICTA]{Influence of bias voltage noise on the Inelastic Cooper-Pair Tunneling Amplifier (ICTA) - Supplementary material}

\author{U. Martel}
\affiliation{ 
Institut Quantique, Université de Sherbrooke, Sherbrooke, Québec, Canada J1K 2R1}
\author{R. Albert}
\author{F. Blanchet}
 \affiliation{ 
Univ. Grenoble Alpes, CEA, INAC-PHELIQS, F-38000 Grenoble, France}
\author{J. Griesmar}
\author{G. Ouellet}
\author{H. Therrien}
\author{N. Nehra}
\author{N. Bourlet}
 \affiliation{ 
Institut Quantique, Université de Sherbrooke, Sherbrooke, Québec, Canada J1K 2R1}
\author{A. Peugeot}
 \affiliation{CNRS, Laboratoire de Physique, Ecole Normale Supérieure de Lyon, Lyon F-69342, France}
\author{M. Hofheinz}
 \email{max.hofheinz@usherbrooke.ca}
 \affiliation{ 
Institut Quantique, Université de Sherbrooke, Sherbrooke, Québec, Canada J1K 2R1}

\date{\today}

\maketitle

\section*{\label{sec:bias_box}Voltage biasing circuits}

\begin{figure}[h]
    \includegraphics[width=0.48\textwidth]{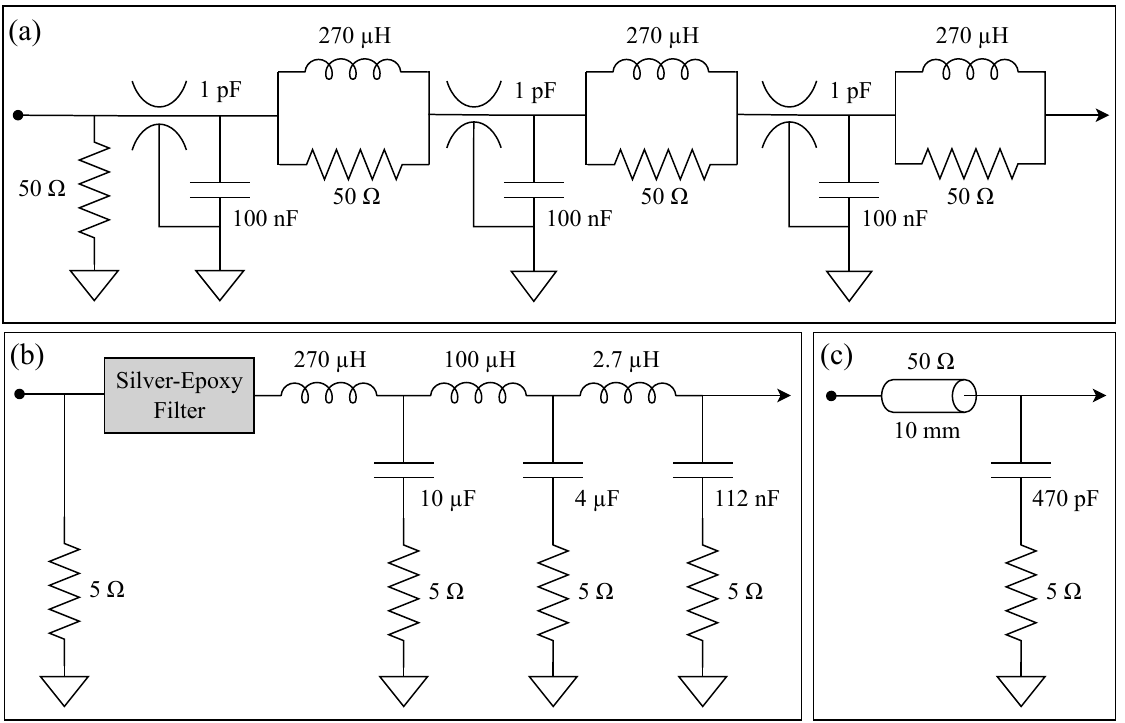}
    \caption{\label{fig:bias_box} Voltage bias circuits schematics. (a) \SI{50}{\ohm} biasing circuit. The 1\,pF capacitors are feed-through capacitors. (b) \SI{5}{\ohm} biasing circuit. (c) Further filtering for Sample A, on the sample holder.}
\end{figure}

Fig.~\ref{fig:bias_box} shows the \SI{50}{\ohm} and \SI{5}{\ohm} biasing circuits used in this article. Both connect to the sample holder using SMA connectors. The biasing resistor in both cases absorbs significant power and heats above the base temperature of the fridge. The subsequent filtering stages filter technical noise as well as thermal noise emitted by the bias resistor and are designed to present a flat output impedance to avoid instabilities of the ICTA. 

The high noise configuration uses the \SI{50}{\ohm} biasing circuit \cite{Grimm2015} illustrated in Fig.~\ref{fig:bias_box}a. Filtering is ensured by three RLC stages in series, with each stage being inside a copper cavity connected via feed-through capacitors to avoid high-frequency leakage. 

The medium and low noise configurations use the \SI{5}{\ohm} biasing circuit \cite{Albert2019} illustrated in Fig.~\ref{fig:bias_box}b. This lower resistance yields less thermal voltage noise than the \SI{50}{\ohm} circuit and the subsequent RLC filter stages have lower cutoff frequency and better high-frequency rejection due a silver epoxy filter stage \cite{Scheller2014}. 


The low noise configuration uses the \SI{5}{\ohm} biasing circuit in conjunction with Sample A, which has further filtering as illustrated in Fig.~\ref{fig:bias_box}c and a final on-chip 100\,pF capacitor. As these filter stages are closer to the actual junctions, they allow a better control of the impedance seen by the junction and prevent parasitic resonances.

\section*{\label{sec:calib}Calibration}

We calibrate gain measurements by measuring the device in a on state at the working point and then divide the measured gain by the gain in an off state where the device acts as a nondissipative linear circuit element. We reach such an off-state by setting the bias voltage far from any operating point (we do not set the voltage to 0 because the SQUID would then enter the superconducting state and act as nonlinear inductor) and by maximally frustrating the SQUID. 

We calibrate noise measurements by first measuring the noise of the device at the working point without any input signal. We then subtract the noise of the subsequent measurement chain obtained by setting the bias voltage to 0 so that the Josephson junction is in the superconducting state where it does not emit noise. This gives the output noise of the amplifier, multiplied by the gain of the subsequent readout chain which we need to calibrate out. 

In a first step we calibrate the gain of the readout chain from the cold microwave switch (see Fig.~1 in the main text) to the digitizer using a Y-factor method, by successively connecting it to a cold and a hot \SI{50}{\ohm} termination. To ensure correct thermalization of these resistances, they are thermally isolated from the switch via superconducting NbTi coax lines and thermally anchored to, respectively, the mixing chamber stage and the still stage of our dilution refrigerator. Their thermal noise is then acquired in the 4 to \SI{12}{\giga\hertz} band. From these two measurements we calculate the gain of the measurement chain from the switches to the digitizer as well as the noise of the amplification chain referred to the microwave switch. 

In a second step we need to calibrate the attenuation of the transmission line between the sample and the microwave switch. To do so, we connect the microwave switch to a short circuit and perform a gain calibration to the microwave switch. 

The ratio of gain calibration to the device and to the switch gives the attenuation between switch and device squared because the signal travels both ways. We, therefore, compute the loss of the transmission line between switch and sample as the square root of this ratio. We observe that this attenuation is flat, indicating that dissipation in the signal mode is negligible and we therefore use a constant attenuation value of 0.498, 0.531 and \SI{0.517}{dB} for, respectively, the low, medium and high noise configuration. It is in good agreement with the expected attenuation of the cable at low temperature. 

The product of this attenuation factor and the gain of the readout chain from switch to digitizer then gives the gain of the readout chain from the sample to the digitizer. Dividing the measured ICTA noise by this gain gives the output noise of the ICTA, which we can then compare to the quantum limit $|\gain|^2-1$.

Note that we use the same readout hardware including the digitizer for both gain and noise measurements, asserting that all measurements are consistent.

\section*{Circuit parameters}

Table \ref{tab:params} shows the circuit parameters for Sample A and Sample B. Sample A is a device which has already been used in Ref\cite{Albert2024}.

Center frequencies $\omegaSI$ and $\widthSI$ are directly measured. $\ZSI$ are estimated from the design. We can then calculate $\phiSI = \sqrt{\pi \frac{4e^2}{h} \ZSI}$ as well as the Josephson energy for which gain diverges ($\Xi = 1$):
\begin{equation}
    \EJ = \frac{\hbar\sqrt{\widthS\widthI}}{\phiS\phiI}.
\end{equation}

\begin{figure}[b]
    \includegraphics[width=0.48\textwidth,clip]{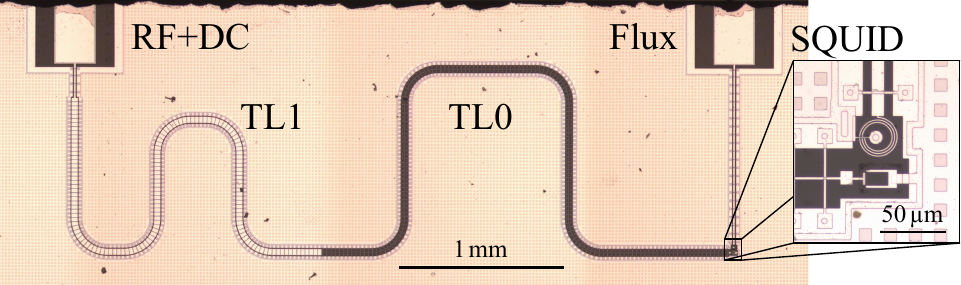}
    \caption{\label{fig:sampleB} Image of a device with same design as sample B. The resonator is formed by a two-segment $\lambda/4$ resonator with expected transmission-line characteristic impedances of approximately \SI{75}{\ohm} (TL0) and \SI{13}{\ohm} (TL1).}
\end{figure}

\begin{table}[h]
    \centering
    \begin{tabular}{c|c|c}
        Parameter & Sample A & Sample B \\
        \hline 
        $\omegaS / 2\pi$ (MHz) & 4800 &  \multirow{2}{*}{4450} \\
        $\omegaI / 2\pi$ (MHz) & 6200 &  \\
        $\widthS/2\pi$ (MHz)& 96 & \multirow{2}{*}{185} \\
        $\widthI/2\pi$ (MHz) & 226 &  \\
        $\ZSI$ (\unit{\ohm})& 400 & 80 \\
        $\phiSI$ & 0.44 & 0.19 \\
        $\EJ/h$ (MHz) @ $\Xi = 1$ & 760 & 5100 \\
    \end{tabular}
    \caption{Key device parameters of both samples. $\omegaSI$ and $\widthSI$ are measured, $\ZSI$ are design parameters, $\phiSI$ and $\EJ$ are calculated from these values.}
    \label{tab:params}
\end{table}

\section*{Image of the device}

An image of sample A can be found in Ref\cite{Albert2024}. Fig.~\ref{fig:sampleB} shows an image of a sample with same design as sample B.




\bibliography{article_biblio}